\documentclass[sigconf]{acmart}
\AtBeginDocument{%
  }

\usepackage{enumitem}
\usepackage{threeparttable}
\usepackage{multirow}
\usepackage{makecell}
\usepackage{tabularx}
\usepackage{siunitx}
\renewcommand\footnotetextcopyrightpermission[1]{}

\begin{document}

\title{SafeGen: LLM-Driven Assertion Generation and Fault Criticality Evaluation for Functional Safety}

\author{
Xuanyi Tan$^{1}$, 
Arjun Chaudhuri$^{1}$, 
Rubin Parekhji$^{2}$, 
Krishnendu Chakrabarty$^{1}$
}

\affiliation{
$^{1}$Arizona State University, Tempe, Arizona, \country{USA}\\
$^{2}$Texas Instruments India, Bengaluru, Karnataka, India
}

\email{
{xtan58, Krishnendu.Chakrabarty}@asu.edu, arjuniitkgp7@gmail.com,  parekhji@ti.com
}

\renewcommand{\shortauthors}{Tan et al.}

\begin{abstract}
With advances in autonomous driving and electric vehicle technologies, functional safety has become a key requirement in automotive chip design. Traditional simulation-based fault analysis tends to be overly conservative at the module level and fails to reflect true fault criticality. 
This paper presents SafeGen, an LLM-driven, formal-verification-assisted framework for functional-safety-oriented fault criticality assessment. 
SafeGen employs large language models (LLMs) with a document-level Hyper Knowledge Graph (HyperKG) that incorporates Failure Modes, Effects, and Diagnostic Analysis (FMEDA) guidelines to extract verifiable specifications from design and safety documents and to evaluate their importance for overall system safety.
The HyperKG is then extended with register-transfer level (RTL) information to guide the generation of Functional Safety Assertions (FSAs) that are semantically grounded and design-aware, each linked to its corresponding specifications for traceable reasoning. 
A gate-to-RTL fault-mapping mechanism supporting both stuck-at and bridging faults, combined with formal property verification (FPV), enables semantic-level fault criticality grading based on specification-linked assertion violations. 
A digital–physical co-simulation platform for a field-oriented control (FOC) system validates SafeGen, demonstrating superior assertion quality compared to other LLM-based assertion generation frameworks. It further provides semantic interpretability in fault criticality assessment when compared to traditional simulation-based approaches.
\end{abstract}

\normalsize
\maketitle

\vspace{-0.8em}
\section{Introduction}
The growing adoption of autonomous driving and electric vehicle technologies continuously increases the complexity of in-vehicle electronic systems. Ensuring safe operation under fault conditions is a formidable challenge. Standards such as ISO 26262 define a unified framework for detecting and managing hardware faults. Their main objective is to ensure that a system maintains a safe operational state even in the presence of faults \cite{palin2011iso, ISO26262-2018, nardi2017functional}.
In the shift-left paradigm, early identification of fault impacts enables timely design-phase countermeasures, thereby improving system reliability \cite{iwata2023new}. To this end, fault simulation and formal methods are employed to evaluate whether a fault can propagate to functional outputs, thus assessing its potential to induce safety risks. These approaches treat fault criticality as a binary classification problem, considering a fault either safe or unsafe depending on whether its effect can be propagated to at least one functional output \cite{CadenceJasperFuSaApp, CadenceXceliumFaultSim, bernardini2016formal}.

At the system level, binary fault classification is reasonable, as faults observable at system outputs typically indicate violations of functional safety \cite{bernardini2016formal}.
However, when applied to the module level, this approach becomes overly pessimistic.
Because modules differ in failure modes and functional roles, similar fault propagation patterns can lead to different risks under varying contexts \cite{cherezova2023understanding, chen2009soc}.
For example, a fault in an auxiliary counter may propagate to the output under specific input stimuli but have little impact on the main control loop. Under the binary classification scheme, such a fault is still labeled as unsafe, thereby triggering the same safety reinforcement measures as truly critical nodes. This conservative strategy results in redundant design, performance loss, and increased testing overhead. Therefore, assessing fault criticality based solely on observability is insufficient, as discussed further in Section ~\ref{sec:fault_eva}. An effective strategy is to evaluate fault significance at multiple levels according to its functional impact through a framework capable of interpreting fault effects from a functional semantics perspective. \par

% Existing studies typically define FSAs as formal assertions used to verify whether safety mechanisms correctly implement their intended safety functions~\cite{}.
% However, in some designs, the system may not include dedicated safety mechanisms or redundant structures. Instead, 
% To accommodate such cases, this work extends the definition of FSAs as follows:
% An FSA is a specialization of functional assertions under the context of functional safety, intended to verify whether critical functions continue to be satisfied under fault assumptions and operational constraints defined by safety analysis.

% A Functional Safety Assertion (FSA) is a functional assertion focused on verifying safety-related functions and behavioral constraints.
% While prior studies define FSAs as formal assertions for checking whether safety mechanisms implement their intended functions~\cite{}, some designs lack explicit safety mechanisms.
% In such cases, safety assurance relies on fault modeling, impact analysis, and safety goal validation.
% Accordingly, we define an FSA as a functional assertion that verifies whether critical functions remain satisfied under fault assumptions and operational constraints defined by safety analysis.

\begin{figure}[t]
    \centering
    \includegraphics[width=0.97\linewidth]{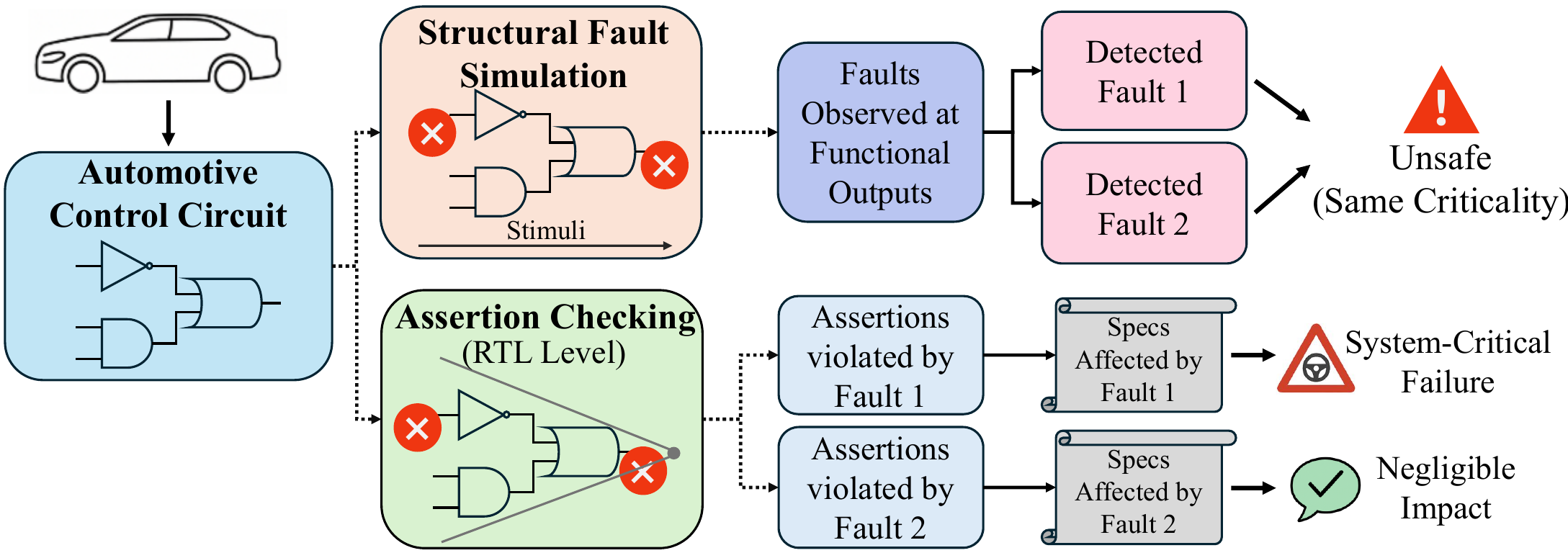}
    \vspace{-1em}
    \caption{The difference between fault simulation and formal methods in evaluating fault criticality.}
    \label{fig:s1}
    \vspace{-1.5em}
\end{figure}

Conventional fault simulation methods based on functional input stimuli are not effective for multi-level fault classification from a functional semantics perspective \cite{bagbaba2022automated, abramovici1988path}. This is because input stimuli rarely map explicitly to circuit functions, and establishing such mappings requires extensive manual effort, making the process impractical for large designs. In contrast, Functional Safety Assertions (FSAs) inherently encode functional semantics. FSAs are assertions designed to verify the robustness and behavioral integrity of a design under fault conditions. Their purpose is to identify faults that lead to functional failures, thereby providing objective evidence for fault classification.
% An FSA can be regarded as a subset of functional assertions that focuses on verifying safety-related functions and behavioral constraints, supporting safety assurance through fault impact analysis and validation of safety goals.
% Each assertion corresponds to a specific module-level function and reflects its correctness under different operating states. 
These assertions are evaluated using Formal Property Verification (FPV), which explores all reachable design states without relying on simulation stimuli. Fig.~\ref{fig:s1} illustrates the difference between fault simulation and FPV in fault criticality evaluation.
Faults are assessed with finer granularity by analyzing assertion violations after fault injection and weighting them based on the importance of the linked specifications. \par

Developing high-quality assertions requires manual effort to interpret specifications, identify safety-related behaviors, and formalize them into precise properties. This process is time-consuming and difficult to scale for complex designs \cite{chao2020evaluating}. Template-based or rule-driven assertion generation frameworks can automate this process, but they struggle to capture high-level semantics and contextual dependencies between design intent and functional behavior \cite{danese2017team, orenes2021autosva}. 
Recent advances in Large Language Models (LLMs) offer new opportunities to address this challenge. With strong semantic reasoning capabilities, LLMs can jointly analyze design documents and register-transfer level (RTL) code to extract design intent and functional semantics, enabling the automatic generation of meaningful SystemVerilog Assertions (SVAs). Recent work has demonstrated the effectiveness of LLMs for generating general and security-related assertions, reducing manual effort and improving coverage \cite{ankireddy2025lasso, kande2024security, yan2025assertllm, bai2025assertionforge, tian2025assertcoder, pulavarthi2025llms, pulavarthi2025assertionbench, paul2025lisa}. For example, AssertionForge builds knowledge-graph-based context retrieval to capture dependencies between specifications and RTL, further enhancing semantic accuracy in assertion generation \cite{bai2025assertionforge, edge2024local}. \par

In this paper, we propose SafeGen, a fault criticality evaluation framework that integrates the semantic reasoning of LLMs with FPV for fine-grained, traceable and interpretable fault criticality assessment.
LLMs use a document-level Hyper Knowledge Graph (HyperKG) to extract verifiable specifications from safety and design documents while assessing their importance to system safety.
The HyperKG is extended in SafeGen with the RTL control data flow graph (CDFG) to guide FSA generation, allowing the LLM to capture global safety and functional semantics and their correspondence to RTL structures.
The generated assertions are linked to the extracted specifications for semantic traceability.
A gate-to-RTL fault mapping is used to analyze gate-level fault effects at the RTL level, supporting both stuck-at and bridging faults.
FPV evaluates injected faults by identifying the violated assertions. With the established mapping between assertions and specifications, each fault is then assigned a criticality based on the importance of the affected specifications.
% enabling traceable and semantically interpretable fault assessment.
The proposed framework is validated on a field-oriented control (FOC) design \cite{WangXuan95_FPGA-FOC_2025} using a digital–physical co-simulation platform and custom functional safety documents that we have developed.
This validation flow demonstrates the effectiveness of the proposed framework in realistic safety-critical scenarios.
% The document-level HyperKG is enhanced through integration with the RTL control data flow graph (CDFG), enabling the LLM to generate FSAs that capture global safety semantics, functional semantics, and their correspondence to local RTL structures.
% The generated assertions are subsequently aligned with the previously extracted specifications.
% A gate-to-RTL fault mapping is used to analyze gate-level fault effects across abstraction levels, supporting both stuck-at and bridging faults.
% Using these assertions, FPV evaluates injected faults by identifying the violated assertions. With the established mapping between assertions and specifications, each fault is then assigned a criticality score based on the importance of the affected specifications, enabling traceable and semantically interpretable fault assessment. The proposed framework is validated on a field-oriented control (FOC) design \cite{WangXuan95_FPGA-FOC_2025} using a digital–physical co-simulation platform and custom functional safety that we have developed.
% This validation flow demonstrates the effectiveness of the proposed framework in realistic safety-critical scenarios.
The proposed approach bridges the abstraction gap between low-level fault analysis and high-level functional semantics, establishing a direct correspondence between properties and gate-level faults. \par
% This approach overcomes the limitations of conventional simulation-based binary fault classification in semantic interpretability. 

The key contributions of this paper are as follows:
\begin{itemize}
\item We propose a novel framework that integrates the semantic reasoning capability of LLMs with FPV to achieve fine-grained fault criticality evaluation at the RTL level.
\item We employ LLMs to extract specifications from design documents and assess their importance to the system, which lays the foundation for traceable and semantically interpretable gate-level fault criticality evaluation.
\item We construct a HyperKG to enhance the LLM’s contextual understanding and enable the generation of high-quality FSAs. To the best of our knowledge, this is the first approach that leverages LLMs to generate FSAs.
\item We extend an existing gate‑to‑RTL fault mapping approach to support analysis of both stuck‑at and bridging faults by mapping gate‑level effects to the RTL.
\item We build a digital–physical co-simulation environment for an FOC-based motor drive system, derive design and functional safety documents, and validate the framework’s effectiveness in a realistic safety-critical scenario.
\end{itemize}

% The remainder of the paper is organized as follows. Section ~\ref{sec:work} describes related prior work. Section ~\ref{sec:fusa} presents the SafeGen framework. Section ~\ref{sec:experiment} describes a case study and an evaluation of SafeGen. Section ~\ref{sec:conclusion} concludes the paper. \par

\section{Related Prior Work}
\label{sec:work}
\subsection{Fault Simulation}

% Simulation-based fault analysis has been extensively studied to estimate fault criticality, but this approach is limited by its binary classification and high computational cost \cite{CadenceXceliumFaultSim}. To accelerate the process, several methods employ neural networks trained on fault simulation data to predict fault criticality \cite{das2024graph, chaudhuri2021fault, chaudhuri2022functional, qutub2022hardware}. Although these methods achieve high inferencing efficiency, they remain dependent on simulation labels and cannot capture functional semantics, which motivates the use of formal verification for semantically grounded fault analysis.
Simulation-based fault analysis has been widely used for fault criticality estimation, but it suffers from binary classification and high computational cost~\cite{CadenceXceliumFaultSim}.
To improve efficiency, recent works employ neural networks trained on fault simulation data~\cite{das2024graph, chaudhuri2021fault, chaudhuri2022functional, qutub2022hardware, chaudhuri2021efficient, chaudhuri2022machine}.
Although these approaches achieve fast inference, they depend on simulation labels and lack awareness of functional semantics.

\subsection{Formal Methods}
% Compared with fault simulation, formal methods can exhaustively explore the input state space without relying on test stimuli, enabling complete coverage of potential faults. Existing formal-method-based fault identification techniques primarily focus on structural detection, where a good machine and a bad machine are constructed and their output differences are analyzed to identify observable faults \cite{traskov2016fault, Cadence_Jasper_FuSa_App, da2020determined}. Such approaches effectively reveal structural defects but fail to evaluate the actual impact of faults on functional behavior.
% Performing formal verification directly on the gate-level netlist provides precise fault localization but greatly increases verification complexity. High-level abstractions are expanded into numerous logic gates, significantly enlarging the cone of influence (COI) and making it difficult for formal engines to converge. To balance scalability and granularity, verification is often performed at the RTL. However, mapping gate-level faults to RTL remains challenging, as internal control and functional signals frequently lack explicit correspondence, making accurate fault injection difficult \cite{iwata2023new}.
% Moreover, structural approaches only determine whether a fault is detectable and cannot distinguish its criticality. In contrast, functional-based formal verification directly evaluates the effect of a fault on functional behavior.

Existing fault identification techniques based on formal verification focus on structural detection; a fault‑free machine and a faulty machine are constructed and their output differences are
analyzed to identify observable faults \cite{traskov2016fault, CadenceJasperFuSaApp, da2020determined, yeung2018whose, da2021automated}.
% These approaches effectively reveal structural defects and can exhaustively explore the input state space without relying on test stimuli, providing full coverage of potential faults. 
However, they remain limited to structural observability and cannot capture the semantic implications of faults on functional behavior, which are essential for functional safety assessment.
Performing formal verification directly on the gate-level netlist provides precise fault localization but increases verification complexity. High-level abstractions (e.g., RTL) are elaborated into a large number of logic gates, significantly expanding the cone of influence (COI) and making it difficult for formal engines to converge \cite{clarke2001bounded}. To mitigate this scalability issue, verification is typically lifted to the RTL abstraction. However, this transition introduces new challenges: the locations of gate-level faults often lack one-to-one signal mappings in RTL, making accurate fault injection difficult. In this work, the impact of gate-level faults is mapped to RTL signals and verified against FSAs, enabling semantically meaningful fault criticality assessment while controlling COI expansion and maintaining convergence. \par

\subsection{LLM-Based Assertion Generation}
LLM-based assertion generation frameworks use multi-LLM collaboration, where subtasks are assigned to specialized LLMs \cite{ankireddy2025lasso, kande2024security, yan2025assertllm, bai2025assertionforge, tian2025assertcoder, pulavarthi2025llms, pulavarthi2025assertionbench, paul2025lisa}. For example, AssertLLM employs multiple LLMs to parse design documents and generate SVAs \cite{yan2025assertllm}. AssertionForge constructs a unified KG integrating design documents with the RTL graph to generate assertions\cite{bai2025assertionforge}. However, it targets module-level port signals while overlooking internal registers and wires that are critical for functional safety. In addition, its simplistic KG representation limits the ability to capture multi-entity interactions and contextual dependencies within design documents.
% AssertCoder targets multimodal inputs and integrates mutation-based evaluation with model checking to form a feedback loop that continuously improves assertion quality \cite{tian2025assertcoder}. 
\subsection{HyperKG}

HyperGraphRAG~\cite{luo2025hypergraphrag} addresses the limitation of graph-based Retrieval Augmented Generation (RAG) methods that model only binary relations. 
It introduces the HyperKG, where hyperedges capture n-ary relations among multiple entities, enabling unified retrieval and generation that improves LLM reasoning accuracy. 
For design functionality and safety documents, such a hypergraph structure is particularly suitable because a module’s safety behavior often depends on the joint interactions among its internal components, such as state machines, communication protocols, and data paths. 
By representing these multi-entity relations as hyperedges that connect internal mechanisms and related signals, the hypergraph allows assertion generation to account for both local structural details and global functional and safety intent. 
In this work, we leverage a functional-safety–aware HyperKG to enhance the LLM’s reasoning capability for generating FSAs.

% HyperGraphRAG \cite{luo2025hypergraphrag} addresses the limitation of graph-based RAG methods that model only binary relations. It introduces HyperKG where hyperedges capture n-ary relations among multiple entities, forming a unified retrieval and generation process that improves LLM accuracy. For design functionality and safety specifications, the hypergraph structure is especially suitable. Consider a safety-critical module that performs control or monitoring functions. Within this module, internal components such as state machines, communication protocols, and data paths jointly determine its safety behavior. When generating assertions, it is not enough to reason about these components separately; their interactions must also be captured to ensure safe operation. A hypergraph can model this multi-entity relationship by representing the module's safety importance as a single hyperedge that connects its internal mechanisms and related signals. 
% This allows the assertion generation process to account for both local structural details and the global functional and safety intent. In this work, we leverage a functional-safety–aware HyperKG to enhance the LLM’s reasoning capability for generating FSAs.
\begin{figure*}[t]
    \centering
    \includegraphics[width=0.8\linewidth]{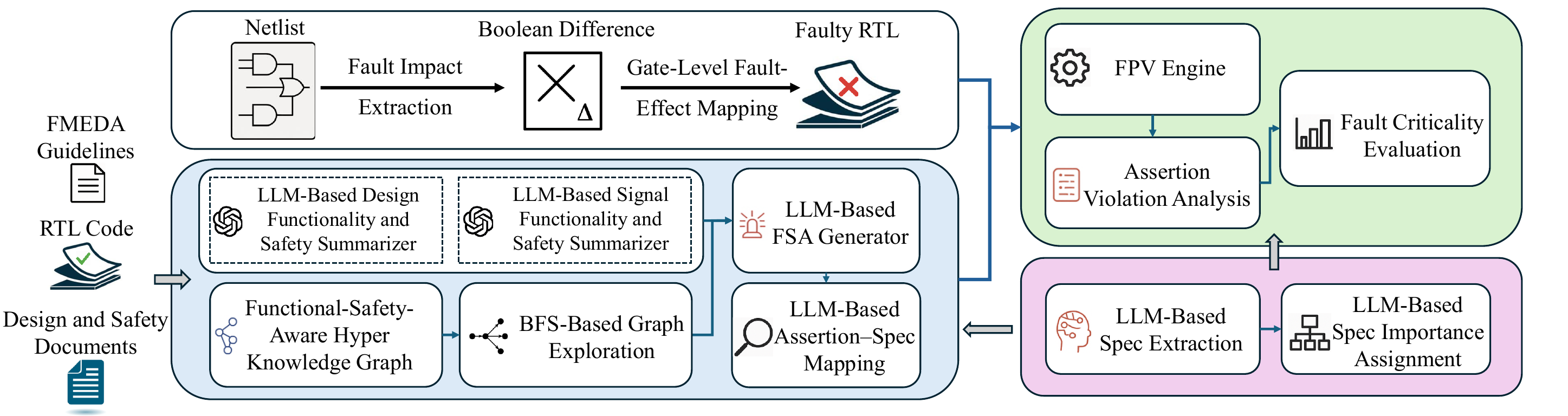}
    \vspace{-1.0em}
    \caption{Assertion generation and gate-level fault criticality evaluation.}
    \label{fig:flow}
    \vspace{-1.0em}
\end{figure*}

\section{Proposed Methodology}
\label{sec:fusa}
\subsection{Overview}
Fig.~\ref{fig:flow} illustrates the framework for multi-level fault criticality assessment. 
We identify and extract specifications from design documents and evaluate their system-level importance. 
FSAs are then automatically generated for RTL signals, and each FSA is linked to its corresponding specifications to establish semantic traceability. 
Next, gate-level fault effects are mapped to the RTL, and FPV is performed on the faulty RTL to evaluate fault criticality. 
% This workflow enables traceable and semantically interpretable evaluation of gate-level fault criticality.

\subsection{LLM-Based Specification Extraction and Importance Assignment}
\label{sec:soec_extract}
% To enable quantitative and semantically interpretable fault criticality assessment, SafeGen employs LLMs to extract specifications from design documents, assign an importance score to each specification, and establish mappings between assertions and their corresponding specifications.
% These mappings allow each fault to be evaluated in terms of the specifications it affects. For instance, when a fault causes multiple assertion violations, the associated specifications can be identified through the mapping, and the 

Since fault criticality depends on the importance of the affected specifications, the consistency, robustness, and reliability of the LLM-generated specifications and their importance scores are crucial to ensure a credible assessment.
Due to the inherent uncertainty of LLM responses, particularly for reasoning models such as GPT-5 whose temperature parameter cannot be tuned, a single response from the LLM may be unreliable \cite{abbasi2024believe}. We use a systematic preprocessing and postprocessing approach to guide the LLM in extracting high-quality, well-structured specifications with stable importance assignments, thereby establishing a trustworthy semantic foundation for fault criticality analysis.

\subsubsection{\textbf{Document-Level HyperKG Construction}}
\label{sec:hkg}
We use HyperGraphRAG as a preprocessing technique to reduce the uncertainty of LLM responses.
The foundation of this technique is our document‑level HyperKG, whose construction begins with LLM‑ based extraction of entities and hyperedges from the Failure Modes, Effects, and Diagnostic Analysis (FMEDA) guidelines, safety documents, and design documents.
% The foundation of this RAG technique is our document-level HyperKG, whose construction of the document-level HyperKG begins with LLM-based extraction of entities and hyperedges from the Failure Modes, Effects, and Diagnostic Analysis (FMEDA) guidelines, safety documents, and design documents.
The entities are defined and extracted according to a custom schema that encompasses functional, structural, and safety-related concepts, such as ports, finite state machines, safe states, and safety levels.
Hyperedges are automatically identified based on the relationships among these entities, capturing their semantic dependencies within the documentation.
This process produces a document-level hypergraph.
All entities and hyperedges are embedded into a high-dimensional vector space to enable vector queries for specification extraction and importance scoring during RAG.
In addition, each entity and hyperedge stores its textual description within attribute fields to provide contextual information for FSA generation.

\subsubsection{\textbf{Specification Extraction}}
During specification extraction, we employ the document-level HyperKG to perform RAG. 
SafeGen automatically extracts all verifiable functional, performance, interface, and safety specifications from safety and design documents. 
The extraction process is designed as a reliable LLM reasoning pipeline that ensures statistical consistency and trustworthiness of the generated specifications. 
To constrain model behavior, we adopt an independent multi‑round inference mechanism in which the identical input configuration is processed by the LLM $N$ times.
Each inference produces a candidate specification set $S_i$ along with a self-evaluated completeness confidence $C_i \in [0,1]$, which reflects how well the model believes its output covers all relevant specifications in the documents. 
The candidate results are parsed into a standard format to form a collection $\{(S_i, C_i)\}$. 
An embedding model is then applied to quantify the semantic similarity among the $N$ generated sets. 
Each $S_i$ is converted into a high-dimensional vector representation, and a cosine similarity matrix $D_s$ is computed. 
The corresponding distance matrix $D = 1 - D_s$ is subjected to agglomerative clustering (average linkage) to identify the most semantically consistent cluster. 
Within this cluster, the sample with the highest confidence $C_i$ is selected as the final specification set.

\subsubsection{\textbf{Importance Assignment of Specifications}}
For the importance assignment of specifications, our objective is to quantitatively evaluate each specification’s contribution to the overall system functionality and safety by leveraging assertions’ capability to verify localized functional behavior and mapping these local properties to system-level functional and safety semantics through semantic reasoning.
To obtain stable and reliable evaluation results, a statistical aggregation approach is employed. 
In the post-processing stage, the LLM performs $M$ independent scoring rounds under identical inputs, producing a complete set of scores $\{I_{i,j}\}$, where $I_{i,j}$ denotes the importance score of the $j$-th specification in the $i$-th inference. 
After $M$ rounds, the final importance score of each specification $j$ is determined as the median of its independently generated scores, i.e., $\tilde{I}_j = \mathrm{median}(I_{1,j}, I_{2,j}, \ldots, I_{M,j})$. 
% The median is chosen because it is insensitive to extreme values and represents the central tendency of the majority judgment, thereby improving interpretability.

\subsection{FSA Generation}
\label{sec:ag}
% We next describe the process of generating FSAs using an LLM enhanced with structured and semantic context. The context is constructed from two complementary sources. First, the HyperKG provides structural and cross-level functional and safety relationships obtained through graph exploration. Second, a set of LLM-based summarizers extracts textual semantics from the functional safety specification and design documentation, capturing intent and constraints related to each signal or module. Together, these sources deliver a comprehensive context that links safety intent, functional behavior, and RTL structure, enabling the LLM to generate assertions that are both semantically consistent and safety-aware.

\subsubsection{\textbf{Document-Level HyperKG Enhancement}}
\label{sec:kg}
% To associate the safety and functional semantics extracted from documentation with the RTL structure and improve the quality of assertion generation, we merge the document-level HyperKG described in Section~\ref{sec:hkg} with the RTL CDFG to form an enhanced HyperKG. The CDFG is extracted from the RTL using PyVerilog \cite{takamaeda2015pyverilog}. The nodes and edges of this graph are annotated with their structural descriptions stored in attribute fields.
% Finally, fuzzy string matching described in \cite{bai2025assertionforge} is applied to align entities between the document-level hypergraph and the RTL CDFG. The matched nodes are then connected to form a unified HyperKG that links functional and safety semantics from the documentation with the structural dependencies of the RTL design. 

\begin{figure}[t]
    \centering
    \includegraphics[width=0.8\linewidth]{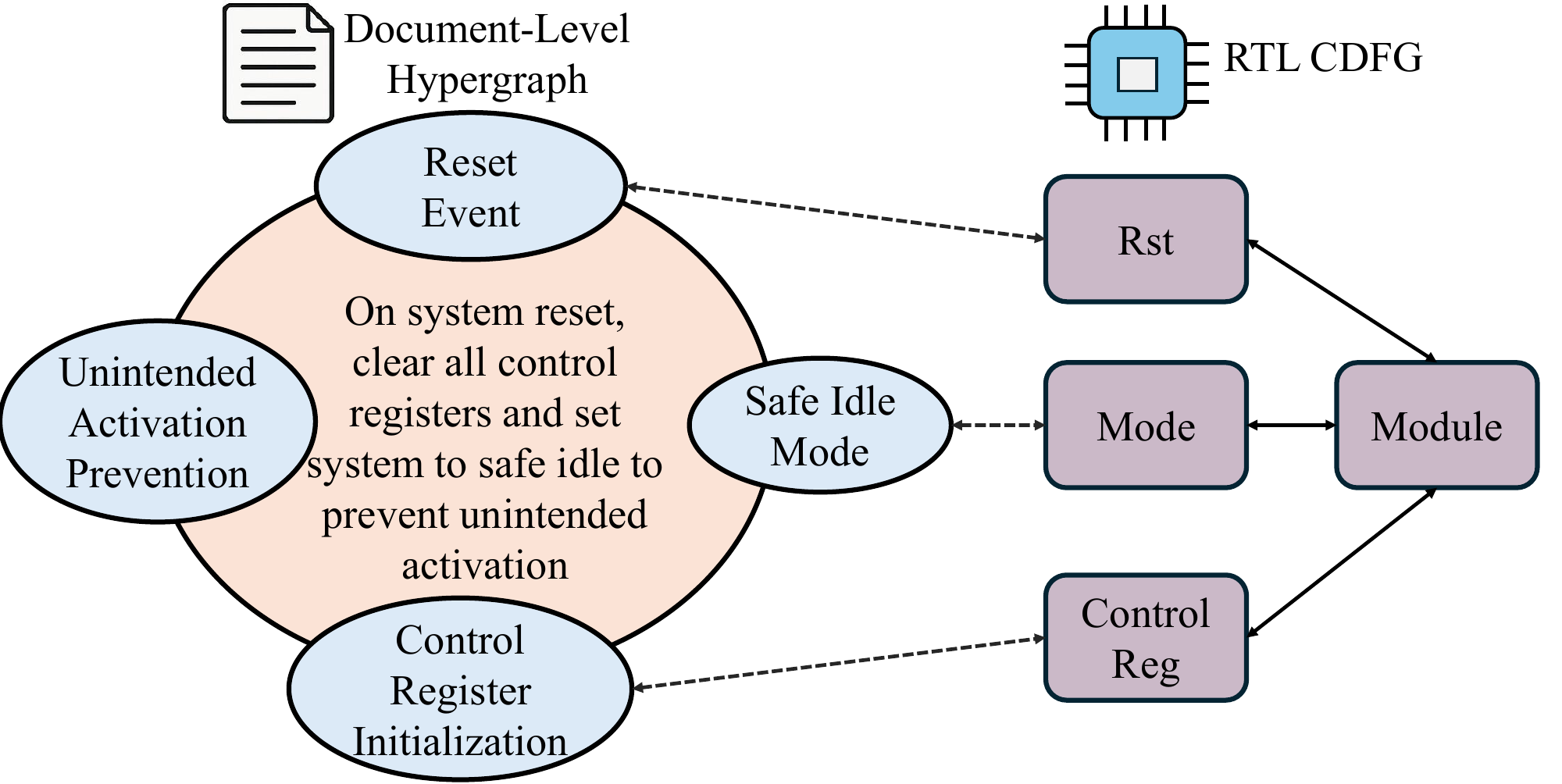}
    \vspace{-1em}
    \caption{Alignment between document HyperKG and RTL CDFG.}
    \label{fig:hyperkg}
    \vspace{-1.5em}
\end{figure}

To associate the safety and functional semantics extracted from documentation with the RTL structure and improve the quality of assertion generation, we merge the document-level HyperKG with the RTL CDFG to form an enhanced HyperKG.
The CDFG is extracted from RTL using PyVerilog~\cite{takamaeda2015pyverilog}, and its nodes and edges are annotated with structural descriptions stored in attribute fields.
Subsequently, fuzzy string matching~\cite{bai2025assertionforge} is applied to align semantically related entities between the document-level hypergraph and the RTL CDFG.
The matched nodes are connected to form a unified HyperKG that links the functional and safety semantics from the documentation with the structural dependencies of the RTL design.
Fig.~\ref{fig:hyperkg} illustrates the alignment between a document-level hypergraph and the RTL CDFG using the reset control logic as an example.
The document states that upon a system reset, all control registers must be cleared and the system must enter a safe idle state.
The LLM extractor identifies key entities such as reset event, register initialization, safe idle mode, and unintended activation prevention, and connects them as a hyperedge representing their causal safety relationship.
The RTL analysis locates the corresponding nodes, including the \texttt{rst}, \texttt{control\_reg}, and \texttt{mode} signals.
These semantic entities and structural nodes are then aligned through fuzzy string matching.

\subsubsection{\textbf{Context Retrieval and Augmentation}}
In SafeGen, FSAs are generated on a per-signal basis.
Before generating an assertion for a given signal, the framework constructs its context from two complementary sources: HyperKG-based graph exploration and multi-LLM summarization.
For the HyperKG exploration, a breadth-first search (BFS) is performed from the node corresponding to the target signal, following a predefined hop limit $h$. 
The search collects all neighboring nodes within $h$ hops together with their associated hyperedges. Textual descriptions from the attributes of these nodes and their corresponding hyperedges are extracted and aggregated to construct the HyperKG-derived context.
In parallel, multiple LLM-based summarizers extract semantic information from FMEDA guidelines, safety documents, and design documents, generating concise descriptions of both design-level and signal-level functionality and safety intent.
Finally, the contexts obtained from HyperKG exploration and LLM summarization are combined to provide a unified, semantically rich context for the LLM, which then generates the corresponding FSAs for each signal.

\subsubsection{\textbf{Assertion–Specification Mapping}}
We use an LLM to identify mappings between assertions and specifications, where a single assertion can be associated with multiple specifications.
This one-to-many relationship enables comprehensive semantic tracing of fault impacts across different functional and safety requirements.

\subsection{Gate-Level Fault Effect Mapping to RTL}

The proposed fault mapping builds upon the approach in \cite{prabhu2012functional}, which computes Boolean differences and performs fault injection at the RTL level using multiplexer-based modeling. However, \cite{prabhu2012functional} mainly focuses on stuck-at faults, whereas functional safety evaluation also requires analyzing bridging faults arising from physical coupling \cite{7586244, jedecJESD88E}. In this work, we extend the above mapping method to support bridging fault analysis.

Typical bridging faults include wired-AND, wired-OR, and dominant bridging faults. To represent bridging fault effects at the RTL level, the abstract bridging fault model must first be converted into a logically equivalent structure, from which a fault-injected netlist can be obtained. For wired-AND (wired-OR) faults, this is achieved by disconnecting the victim and aggressor nets and feeding them into an AND (OR) gate, whose output drives all fan-out nodes of the two original nets. For dominant bridging faults, the victim net is disconnected, and the aggressor net directly drives all fan-out nodes originally driven by the victim.

\begin{figure}[t]
    \centering
    \includegraphics[width=0.65\linewidth]{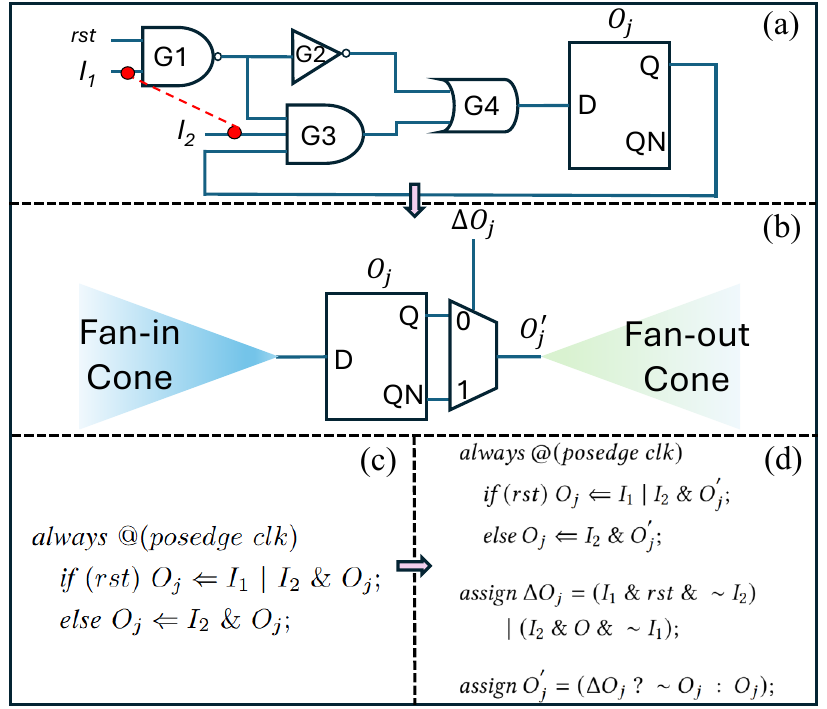}
    \vspace{-1em}
    \caption{Mapping of wired-AND bridging fault effects from the netlist to RTL.
(a) Netlist with injected fault.
(b) Insertion of bridging fault effect at the RTL level.
(c) Fault-free RTL code.
(d) RTL code with inserted bridging fault effect.}
    \label{fig:mapping}
    \vspace{-1.8em}
\end{figure}

For each bridging fault, the set of primary outputs or registers within the combined fan-out cone of the aggressor and victim nets is defined as $O = \{O_1,\allowbreak O_2,\allowbreak \dots,\allowbreak O_j\}$, where each $O_j$ in the gate-level netlist directly corresponds to an RTL primary output or register. For each $O_j \in O$, the corresponding fan-in cone is extracted from both the fault-free and faulty netlists, yielding two Boolean functions $O_{j}^{\text{ff}} = f^{\text{ff}}(I_1^{\text{ff}}, I_2^{\text{ff}}, \dots, I_m^{\text{ff}})$ and $O_{j}^{\text{ft}} = f^{\text{ft}}(I_1^{\text{ft}}, I_2^{\text{ft}}, \dots, I_n^{\text{ft}})$, where $I_m^{\text{ff}}$ and $I_n^{\text{ft}}$ denote the primary inputs or registers in the fan-in cone of $O_j$ in the fault-free and faulty netlists, respectively. Because a bridging fault alters signal dependencies, the two Boolean functions $f^{\text{ff}}(\cdot)$ and $f^{\text{ft}}(\cdot)$ are not logically identical.
The Boolean difference between the fault-free and faulty outputs is defined as $\Delta O_j = O_{j}^{\text{ff}} \oplus O_{j}^{\text{ft}}$. When $\Delta O_j = 1$, a logical deviation occurs at the output $O_j$, indicating that the fault effect is activated. When $\Delta O_j = 0$, the output behavior remains identical to the fault-free case. After deriving the Boolean difference expression of $\Delta O_j$, the fault effect can be mapped to the RTL to construct the faulty design. Fig. ~\ref{fig:mapping} illustrates the mapping of a wired-AND fault from the gate-level netlist to the RTL. This mapping is realized by inserting a multiplexer (MUX) at the corresponding RTL primary output or register. The MUX models the activation of the bridging fault: its select signal is driven by $\Delta O_j$, which determines whether the fault effect is activated or the normal signal is preserved. 
The MUX output $O_{j}^{'}$ represents the RTL signal that reflects both the fault-free and faulty behaviors, depending on the selection controlled by $\Delta O_j$.

To maintain logical consistency, signal references in the RTL must be updated accordingly: all occurrences of $O_j$ on the right-hand side are replaced with $O_{j}^{'}$. Through this process, the activation and propagation behavior of bridging faults can be accurately modeled at the RTL level, providing a foundation for subsequent formal verification and fault criticality analysis.

% For each low-criticality or undetected fault $f_i$, a structural analysis based on the COI is performed on the gate-level netlist to identify the primary outputs and registers that can be structurally affected over multiple time frames. The resulting fault impact set $F_i$ comprises these affected outputs and registers.

% Once $F_i$ is obtained, the framework can identify which signals may potentially be influenced by fault $f_i$, but it remains unclear whether these influences are functionally safety-relevant. To address this, a LLM-based fault classifier is developed based on the functional-safety-aware knowledge graph constructed in \cref{sec:ag}. The classifier queries the knowledge graph using the signals in $F_i$ to determine whether they are associated, directly or indirectly, with any safety-related concepts defined in the functional safety specification. If the knowledge graph query indicates that none of the signals in $F_i$ are mapped to any safety-related nodes, fault $f_i$ is classified as benign. Conversely, if at least one signal in $F_i$ is linked to a safety concept, fault $f_i$ is categorized as potentially unsafe. 

% For such faults, the associated signals are fed back into the functional-safety-oriented AssertionForge to generate new test plans and assertions. 

\section{Case Study}

\label{sec:experiment}
\subsection{Experimental Setup}
In our study, all stuck-at faults are extracted from the netlist using its graph representation, and equivalent faults are collapsed. For bridging faults, the physical layout is extracted using Synopsys IC Compiler, and coupling capacitances are obtained from Synopsys StarRC reports. The fault list is generated by Synopsys TestMAX. Bridging faults that cause combinational logic loops are removed. FPV is performed with Cadence JasperGold, while fault simulation for baseline comparison is conducted with Cadence Xcelium. 
% In this experiment, the SafeGen’s fault criticality assessment is compared against traditional fault-simulation-based analysis to evaluate accuracy. 
Fault simulation and FPV are conducted on a workstation with 32 physical cores (64 threads) using AMD EPYC 7313 processors. The hop limit $h$ for the BFS-based HyperKG exploration is set to 5.
All LLM-based experiments are conducted on GPT-5.

\subsection{Design and Document Preparation}
\subsubsection{\textbf{Setup Checklist}}
In this work, an open-source FOC design is adopted as the evaluation target \cite{WangXuan95_FPGA-FOC_2025}. Since the repository provides only the RTL code, we develop a physical motor model along with safety and design documentation to enable closed-loop motor simulation and safety-critical scenario analysis.

\subsubsection{\textbf{Design Overview}}
\begin{figure}[t]
    \centering
    \includegraphics[width=0.91\linewidth]{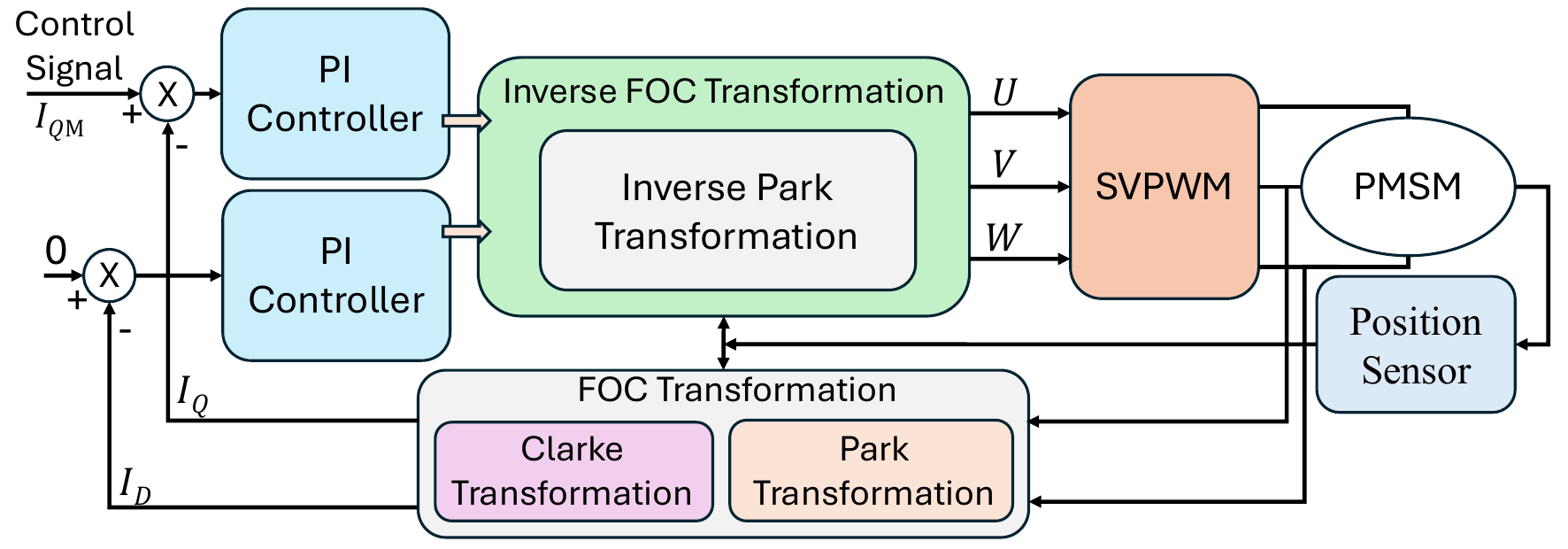}
    \vspace{-1em}
    \caption{FOC design for driving a PMSM \cite{WangXuan95_FPGA-FOC_2025}.}
    \label{fig:foc}
    \vspace{-0.5em}
\end{figure}

The system adopts an FOC-based control method to drive a Permanent Magnet Synchronous Motor (PMSM); see Fig. ~\ref{fig:foc}.
It consists of a digital control subsystem and a physical motor subsystem.
The digital subsystem acquires current and rotor-angle signals, performs Clarke and Park transformations, executes Proportional–Integral (PI) regulation, and generates three-phase Pulse-Width Modulation (PWM) signals via Space-Vector PWM (SVPWM) \cite{merzoug2008comparison, kontarvcek2015cost}.
The physical subsystem, represented by the PMSM, receives these PWM signals and provides current and position feedback, forming a closed-loop current control that ensures stable torque regulation.

% The system also supports flexible fault injection to evaluate the controller’s response and fault tolerance under abnormal conditions. Because the design accurately emulates typical motor operating scenarios in automotive applications, it serves as an experimental platform for functional-safety analysis and fault-behavior characterization.

\subsubsection{\textbf{Modeling and Simulation}}
\begin{figure}[t]
    \centering
    \includegraphics[width=0.6\linewidth]{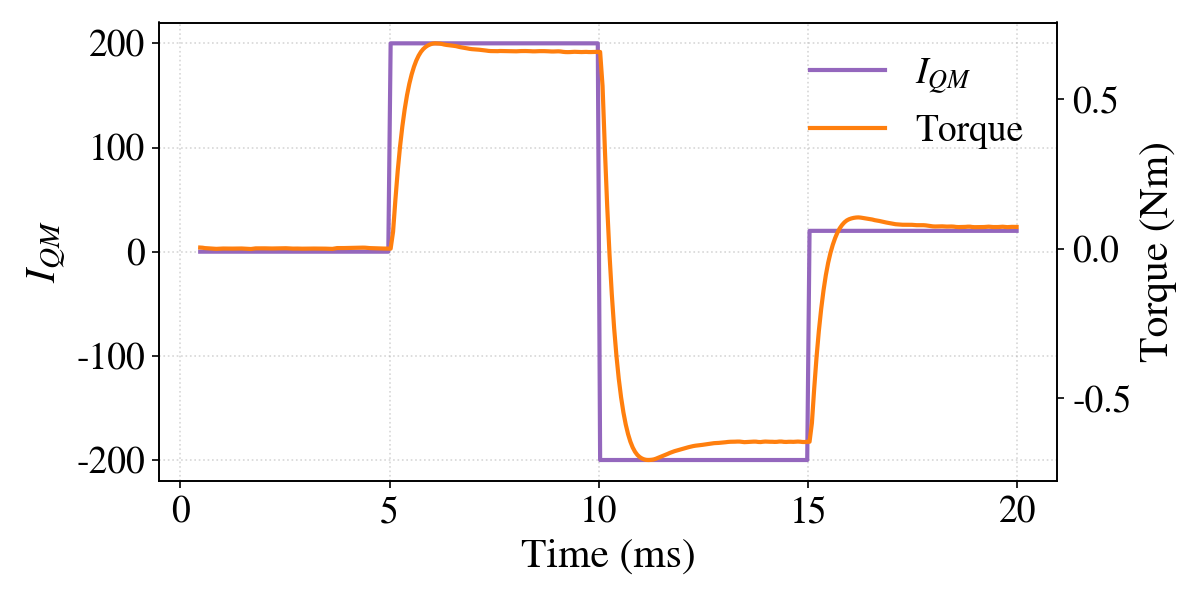}
    \vspace{-1.5em}
    \caption{Closed-loop simulation of the FOC system.}
    \vspace{-1.7em}
    \label{fig:torque}
\end{figure}

% Since the open-source FOC design provides only the RTL code of the digital control system, we construct a physical model of the PMSM to enable closed-loop control. 
The physical model includes an inverter model, a PMSM electrical model, a mechanical dynamics model, and a position sensor model. 
The three-phase PMSM has seven pole pairs and is driven by a 24~V DC bus.
Its key electrical and mechanical parameters are as follows: phase resistance 0.5~$\Omega$, inductance 0.4~mH, flux linkage 0.03~Wb$\cdot$turn, rotor inertia 0.002~kg$\cdot$m$^2$, viscous friction 0.0005~N$\cdot$m$\cdot$s, and load torque 0.02~N$\cdot$m.
The system employs an $I_D = 0$ FOC strategy, using the quadrature-axis current $I_{Q\mathrm{M}}$ to regulate motor torque.
% The PI regulator and FOC algorithm generate SVPWM signals to drive the inverter, enabling stable torque output.
Figure~\ref{fig:torque} shows the current and torque waveforms from the closed-loop simulation.
When $I_{Q\mathrm{M}} > 0$, the motor produces positive torque for propulsion; when $I_{Q\mathrm{M}} < 0$, it generates negative torque for braking.
The results indicate good dynamic response and stable steady-state performance.

% By tuning the model parameters, the system achieved accurate torque control within the closed-loop configuration. The physical model includes an inverter model, a PMSM electrical model, a mechanical dynamics model, and a position sensor model. 
% The entire system allows the motor torque to be controlled through a user-defined reference current $I_{QM}$. The simulation results are shown in \cref{fig:torque}, which demonstrate that the closed-loop system achieves precise and stable torque control.

\subsubsection{\textbf{Documentation Preparation}}
\label{sec:design}

% \begin{table}[t]
% \centering
% \caption{Module safety classification based on fault simulation results.}
% \renewcommand{\arraystretch}{1.1} % row
% \setlength{\tabcolsep}{13pt}      % col
% \begin{tabular}{|c|c|c|}
% \hline
% \textbf{Module} & \textbf{$M_f$} & \textbf{Safety Level} \\
% \hline
% clark\_tr          & 203 / 294 & D \\
% \hline
% hold\_detect       & 23 / 38   & D \\
% \hline
% svpwm              & 260 / 446 & D \\
% \hline
% foc\_top           & 145 / 260 & D \\
% \hline
% sincos             & 93 / 210  & C \\
% \hline
% pi\_controller\_iq & 141 / 444 & B \\
% \hline
% park\_tr           & 101 / 324 & B \\
% \hline
% cartesian2polar    & 113 / 440 & A \\
% \hline
% pi\_controller\_id & 87 / 444  & A \\
% \hline
% \end{tabular}
% \label{tab:module_class}
% \vspace{-2em}
% \end{table}

The developed design document describes the overall system functionality, module operations, interfaces, state machines, and other design details.
To derive the functional safety document, an RTL-level stuck-at fault simulation is performed.
The impact of each injected fault is evaluated by monitoring the motor torque variation. 
When the torque deviation exceeds $\pm 10\%$ of its nominal value, the fault is considered critical.
This threshold is derived from the functional behavior of the system and serves as the safety objective for this analysis. 
All injected faults are classified into two categories according to their impact on the system:
(1) safe faults, which do not cause observable torque deviations, and 
(2) unsafe faults, which lead to deviations beyond the $\pm 10\%$ threshold. 
Such fault simulations enable quantitative evaluation of the design's fault vulnerability and reveal which registers or logic modules have the most significant influence on torque stability. 
All of this information is included in the functional safety document, which provides the LLM with the semantic and structural context for extracting safety requirements and generating FSAs. Note that the RTL-level fault simulation is used only to construct the functional safety documents and is not involved in subsequent gate-level fault criticality assessment.

% In this study, register-level stuck-at fault simulations are conducted by sequentially injecting stuck-at faults into each register in the RTL design and observing the system's response to each fault. 

% For the functional safety specifications, RTL-level fault simulation is employed to analyze fault tolerance and determine the safety level of each module.

% Specifically, single stuck-at faults are injected into all signals within the RTL modules, and closed-loop fault simulations are conducted to evaluate how each fault affects torque control performance. A fault is considered critical if, under the fault, the torque deviates by more than $\pm 10\%$ from its nominal value. For each module, we calculate the ratio $M_f = \frac{\text{faults causing system failure}}{\text{total injected faults}}$, which quantifies the module’s contribution to overall system safety. Modules are then classified into safety levels (D/C/B/A) according to their $M_f$ values, as shown in ~\ref{tab:module_class}. For high-criticality modules, module-level safety goals and safe states are defined. All of this information is integrated into the developed functional safety specifications for FSA generation.

\begin{table*}[t]
\centering
\caption{Assertion quality evaluation (SafeGen versus prior solutions).}
\vspace{-3.5mm}
\small
\renewcommand{\arraystretch}{0.7}
\setlength{\tabcolsep}{10pt}
\begin{tabular}{|c|c|c|c|c|c|c|c|c|}
    \hline
    \multirow{2}{*}{\textbf{Model}} 
    & \multirow{2}{*}{\textbf{\#SVA}} 
    & \multirow{2}{*}{\textbf{\#SynC}} 
    & \multirow{2}{*}{\textbf{\#Proven}} 
    & \multicolumn{4}{c|}{\textbf{COI Coverage (\%)}} \\ 
    \cline{5-8}
    & & & 
    & \textbf{Statement} & \textbf{Branch} & \textbf{Functional} & \textbf{Toggle} \\
    \hline
    AssertLLM \cite{yan2025assertllm}  
    & 424 & 313 & 83 & 66.59 & 66.99 & 65.05 & 56.20 \\

    AssertionForge \cite{bai2025assertionforge}  
    & 667 & 460 & 105 & 66.75 & 67.02 & 66.13 & 62.31 \\

    SafeGen  
    & 976 & 745 & 250 & 83.68 & 83.73 & 83.83 & 84.48 \\

    SafeGen w/o HyperKG  
    & 625 & 459 & 170 & 66.55 & 66.96 & 64.53 & 53.25 \\

    SafeGen for fault evaluation  
    & 8541 & 7243 & 1748 & 100.00 & 100.00 & 99.93 & 99.60 \\
    \hline
\end{tabular}
\vspace{-0.5em}
\label{tab:m_metric_calculation}
\end{table*}

\subsection{Assertion Quality Evaluation}

To evaluate the quality of FSAs generated by our framework, we compare its performance with AssertLLM and AssertionForge~\cite{bai2025assertionforge, yan2025assertllm}.
For a fair comparison, the prompts of both baselines are adjusted to emphasize functional safety.
Since AssertionForge supports only module port signals, the comparison is limited to those signals.
We use multiple evaluation metrics summarized in Table~\ref{tab:m_metric_calculation}.
Cadence JasperGold is employed to measure the total number of generated assertions (\#SVA), syntactically correct assertions (\#SynC) and formally proven assertions (\#Proven), excluding vacuous proofs.
COI coverage of the proven assertions is further analyzed using JasperGold’s COI-based models (Functional, Branch, Statement, and Toggle).
An ablation study examines the effect of removing the HyperKG on SafeGen’s assertion quality.
We also present the assertions used for fault evaluation, which are generated not only for the module ports but also for all internal signals within each module.
Assertions that are difficult to converge are removed from the evaluation set.
The results show that SafeGen outperforms both baselines in assertion quality, and removing the HyperKG leads to a noticeable degradation in performance.

\subsection{LLM Reliability in Specification Extraction and Importance Assignment}

Following the procedure in Section ~\ref{sec:soec_extract}, the reasoning pipeline is applied to the same design document for 15 independent runs ($N=15$), each producing a specification set $S_i$ with confidence $C_i$.
Cosine similarity among all $S_i$ is computed, and agglomerative clustering (average linkage, threshold = 0.05) groups samples with similarity above 0.95.
This process yields four clusters of sizes 10, 2, 2, and 1. 
The dominant cluster (10 samples) demonstrates strong convergence toward a stable specification representation. 
The sample with the highest $C_i$ in this cluster yields 70 final specifications for downstream fault criticality analysis.
The importance score of each specification ranges from 0 (negligible impact) to 4 (system-critical failure).
To ensure stability, the LLM performs $M=30$ independent scoring rounds, and the final importance score of each specification is taken as the median across runs.
We observe all specification importance scores have standard deviations below 1, indicating high consistency across repeated evaluations.

\subsection{Gate-Level Fault Criticality Analysis}
\subsubsection{\textbf{Fault Criticality Definition}}
% To evaluate the fault criticality, we perform FPV on the gate-level fault-effect injected RTL using the proven assertions. 
% During FPV, we record the set of violated assertions $\mathcal{A}_f$ for each fault $f$. 
% Each violated assertion in $\mathcal{A}_f$ belongs to one of two categories: (1) a vacuous proof, where a previously proven assertion becomes vacuously true, and (2) a counterexample violation, where FPV identifies a counterexample for a previously proven assertion. 
% Based on the predefined mapping between assertions and specifications, each violated assertion $a \in \mathcal{A}_f$ corresponds to a specification set $\mathcal{S}_a$, and the overall set of affected specifications for fault $f$ is given by $\mathcal{S}_f = \bigcup_{a \in \mathcal{A}_f} \mathcal{S}_a$. 
% Each specification $s_j \in \mathcal{S}_f$ has an importance score $\tilde{I}_j$ assigned during the LLM-based importance assignment stage, and the fault criticality score $C_f$ is calculated as $C_f = \sum_{s_j \in \mathcal{S}_f} \tilde{I}_j$. 

% To evaluate fault criticality, we perform FPV on the gate-level fault-effect injected RTL using the proven assertions. 

To evaluate the fault criticality, we perform FPV on the gate-level fault-effect injected RTL using the proven assertions.
For each fault $f$, FPV records the violated assertion set $\mathcal{A}_f$, where each violation is either (1) a vacuous proof, meaning a previously proven assertion becomes vacuously true, or (2) a counterexample violation, where FPV finds a counterexample to a proven assertion. 
Based on the predefined mapping between assertions and specifications, each violated assertion $a \in \mathcal{A}_f$ corresponds to a specification subset $\mathcal{S}_a$, and the affected specification set is $\mathcal{S}_f = \bigcup_{a \in \mathcal{A}_f} \mathcal{S}_a$. 
As each specification $s_j \in \mathcal{S}_f$ carries an importance score $\tilde{I}_j$, the fault criticality is computed as $C_f = \sum_{s_j \in \mathcal{S}_f} \tilde{I}_j$.

\vspace{-0.5em}

\subsubsection{\textbf{Comparison with Simulation-Based Methods}}
\label{sec:fault_eva}

\begin{table}[t]
\centering
\small
\caption{Number of faults detected by different methods. Values in parentheses: number of faults detected by both SafeGen and the corresponding method.}
\vspace{-3.5mm}
\renewcommand{\arraystretch}{0.7}
\setlength{\tabcolsep}{1pt}
\begin{tabular}{|c|c|c|c|c|}
    \hline
    \multirow{3}{*}{\textbf{Method}} 
    & \multicolumn{2}{c|}{\textbf{Bridging faults}}
    & \multicolumn{2}{c|}{\textbf{Stuck-at faults}} \\ 
    \cline{2-5}
    & \textbf{\# Detected} 
    & \makecell{\textbf{\# Total}\\\textbf{collapsed}}
    & \textbf{\# Detected} 
    & \makecell{\textbf{\# Total}\\\textbf{collapsed}} \\
    \hline
    Structural simulation & 4688 (1767)* &  & 39240 (12441)* &  \\
    Functional simulation & 794 (656) & 16495 & 4889 (3940) & 183346 \\
    SafeGen        & 2287       &  & 18869 &  \\
    \hline
\end{tabular}
\vspace{1mm}
\begin{minipage}{0.95\linewidth}
\raggedright
\footnotesize{*The comparison between structural fault simulation and other methods shows that it is overly pessimistic.}
\end{minipage}
\vspace{-2em}
\label{tab:detect}
\end{table}

To evaluate SafeGen, we employ two comparative setups. 
The first setup is the structural fault simulation baseline, in which the netlist of the digital subsystem is simulated under open-loop functional stimuli. 
A fault is labeled unsafe if its effect propagates to outputs.
This approach represents a typical method for fault grading, focusing solely on whether a fault can be propagated to observable outputs.
The second setup is functional fault simulation, which serves as a ground truth reference. 
It is performed in a closed-loop configuration, where control input $I_{QM}$ is adjusted to regulate motor current. 
During simulation, we monitor whether the motor torque deviation under faults exceeds $\pm 10\%$ of the nominal value. 
Faults causing deviations beyond this threshold are classified as safety-critical, as they lead to system errors. 
While functional fault simulation provides accurate assessment of fault criticality by capturing real system behavior, it cannot serve as a baseline. This is because safety verification is performed before system integration and the selection of physical components such as the motor and load, making closed-loop functional fault simulation infeasible at the stage when fault grading is most needed.
Therefore, we employ functional fault simulation as a reference to: (a) validate SafeGen's detection accuracy, and
(b) quantify the pessimism of structural fault simulation, which tends to conservatively label faults as unsafe.
\par

% for two reasons:
% (1) It requires physical modeling (e.g., motor modeling, load characteristics) that is typically unavailable to digital hardware designers during the design phase.
% (2) Safety verification occurs often before system integration and physical component selection, making closed-loop functional simulation infeasible at the stage when fault grading is most needed.

% The functional fault simulation is used for two purposes: 
% (1) to validate the fault detection accuracy of SafeGen, and 
% (2) to provide a reference for analyzing the pessimism of structural fault simulation.
% Functional fault simulation cannot be directly applied to fault criticality assessment in practical scenarios, as such evaluation typically depends on specific downstream applications (e.g., load or physical modeling), which lies beyond the scope of digital hardware designers.
% Fundamentally, there is an essential difference in the verification environment: functional fault simulation operates in a closed-loop environment, whereas both structural fault simulation and SafeGen evaluate faults within an open-loop digital subsystem. Therefore, we employ functional fault simulation as a reference, due to it can reflect real system behavior under fault conditions. \par

Table~\ref{tab:detect} lists the number of detected bridging and stuck-at faults across the three methods. 
The fault coverage is relatively low because the simulations are driven by functional, rather than test patterns. Note that faults not included in the ``detected'' list in Table~\ref{tab:detect} are not important for functional safety. Unlike dedicated fault-detection assertions, SafeGen employs FSAs to evaluate the criticality of faults rather than to maximize coverage. SafeGen detects 80.6\% of the stuck-at faults and 82.6\% of the bridging faults detected by the functional fault simulation, demonstrating that its assertions effectively capture faults that lead to system errors.
Notably, 3894 bridging faults and 34351 stuck-at faults detected by structural fault simulation did not result in any system errors in the functional simulation.
As a result, although many faults are detected at the outputs of the digital subsystem and labeled as unsafe, most have no impact on the torque control of the FOC-based motor drive system, indicating that the structural fault simulation is overly pessimistic in assessing fault criticality.

\begin{figure}[t]
    \centering
    \includegraphics[width=1\linewidth]{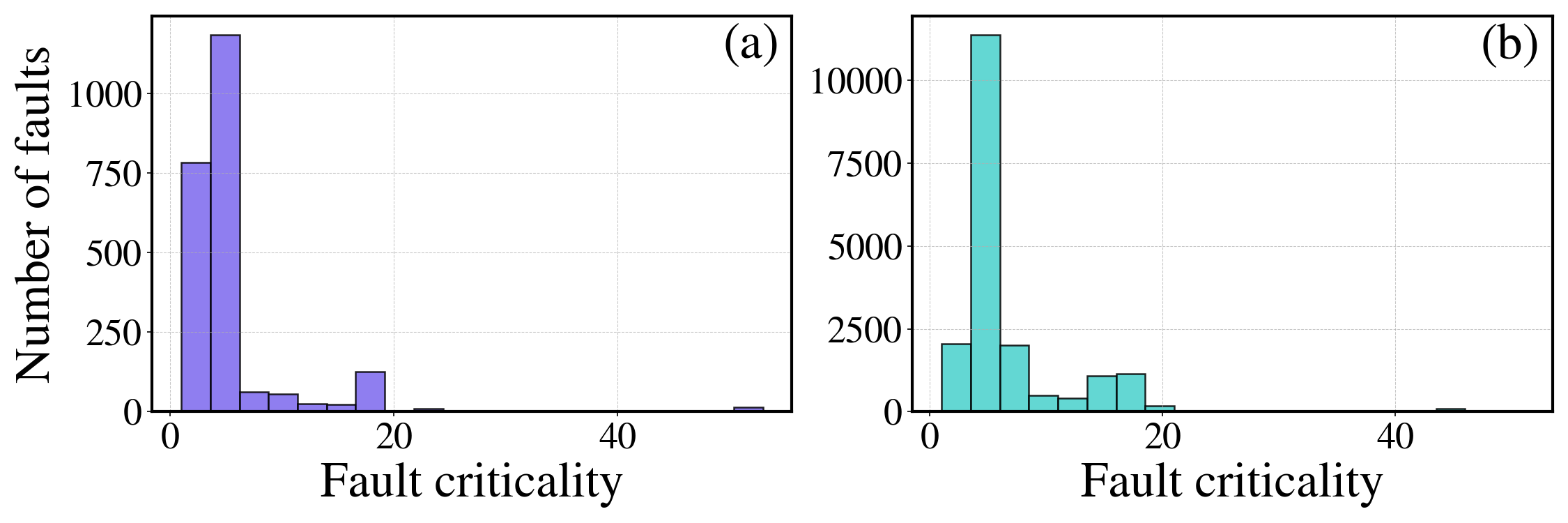}
    \vspace{-1.5em}
    \caption{Fault criticality distribution of bridging faults (a) and stuck-at faults (b).}
    \label{fig:fcd}
    \vspace{-1.5em}
\end{figure}

Fig.~\ref{fig:fcd} presents the fault criticality distribution evaluated by SafeGen.
Among the highly critical faults, most are related to reset correctness, PI numeric saturation, and control arithmetic operations.
These components are essential for maintaining the stability of the closed-loop motor control system, and faults in these regions can destabilize the current loop and lead to unsafe voltage levels. In contrast, low-criticality faults mainly affect timing or observability without significantly influencing system stability.
As expected, the majority of faults exhibit low criticality; thus SafeGen effectively addresses the difficult problem of identifying the "needle in the haystack".
Moreover, each fault’s criticality can be directly traced to the specifications it violates, providing engineers with clear guidance for design refinement and functional safety improvement.

\subsection{Efficiency Analysis}
We compare the runtime between the structural fault simulation baseline and the SafeGen for stuck-at faults.
The runtime of fault simulation consists of two parts: preparation of the testbench, which is a manual task and thus not measurable in execution time, and the fault simulation itself, which takes 5.5 days to complete.
For SafeGen, the construction of the HyperKG takes 2 minutes, specification extraction and scoring take 20 minutes, assertion generation takes 12.5 hours, and gate-to-RTL fault mapping takes 6.7 hours.
The FPV-based fault evaluation takes 3.1 days.
The fault-to-RTL mapping and other LLM-based processes can be executed in parallel, therefore the overall runtime is 3.7 days. Overall, SafeGen achieves considerably better efficiency than fault simulation.

\section{Conclusion}
\label{sec:conclusion}
We have presented an LLM-driven, formal-verification-assisted framework for fault criticality assessment for functional safety.
We constructed a HyperKG to organize the extracted specifications, assess their importance, and support the generation of high-quality FSAs through enhanced understanding of design and safety semantics.
We have integrated gate-to-RTL fault mapping with FPV to bridge the gap between low-level fault analysis and high-level functional semantics.
We have developed a digital–physical co-simulation environment for an FOC system and derived functional safety and design documents to validate our framework. Results show that SafeGen generates higher-quality assertions than existing LLM-based frameworks. Its semantic-level fault criticality assessment achieves high accuracy, with each fault’s criticality traceable to specific functional and safety specifications.
% We demonstrate that our approach achieves better semantic interpretability than traditional simulation-based methods. Future work explores mapping the effects of gate-level delay faults, including transition faults, to RTL so that the static behavior of each gate-level node can be assessed comprehensively.

\begin{acks}
This research was supported in part by the Semiconductor Research Corporation under Contract No. 3177.001.
\end{acks}

\balance
\bibliographystyle{ACM-Reference-Format}
\bibliography{main}

\end{document}